\documentclass[amsmath,11pt]{article}

\usepackage{amssymb,amssymb,epsfig,bm}
\usepackage{caption}
\usepackage{subcaption}

\date{\empty}

\begin{document}

\title{\bf The deceleration parameter in perturbed\\ Bianchi universes with a peculiar-velocity ``tilt''}
\author{Amalia Tzartinoglou${}^1$ and Christos G. Tsagas${}^{1,2}$\\ {\small ${}^1$Section of Astrophysics, Astronomy and Mechanics, Department of Physics}\\ {\small Aristotle
University of Thessaloniki, Thessaloniki 54124, Greece}\\ {\small ${}^2$Clare Hall, University of Cambridge, Herschel Road, Cambridge CB3 9AL, UK}}

\maketitle

\begin{abstract}
Bianchi cosmologies are ``natural'' anisotropic extensions of the Friedmann universes and they have long been used to investigate the cosmological implications of anisotropy. The latter introduces new ingredients to the standard scenarios, although there are physical processes and effects that maintain their basic Friedmann features when extended to Bianchi universes. Here, we assume a perturbed Bianchi model and look into the implications of the observers' peculiar flow for their measurement and their interpretation of the deceleration parameter. Our motivation is twofold. To begin with, relative motions have long been known to deceive the observers by ``contaminating'' the observations, which also still suffer from sample limitations that cloud the statistical significance of the findings. Further motivation comes from claims that observers in bulk flows that expand slightly slower than their surroundings can have the illusion of cosmic acceleration in a universe that is actually decelerating. The claim was originally based on studies of a perturbed tilted Einstein-de Sitter model, but persisted when the background cosmology was replaced by any of the three Friedmann universes. This raised the possibility that the peculiar-motion effect on the deceleration parameter may be generic and largely independent of the host spacetime. Here, we investigate this possibility by extending the earlier studies to perturbed Bianchi models. We find that the Friedmann picture remains unchanged, unless the Bianchi background has unrealistically high anisotropy. The bulk-flow observers can still be misled to the illusion of accelerated expansion by their own peculiar motion.
\end{abstract}

\section{Introduction}\label{sI}
We are all familiar with some of the illusions relative motion can create. Passengers on a train that slows down, for example, may believe for a moment that it is the train next to theirs which has accelerated away (and vice versa). It is not only in our everyday life, however, that such misunderstandings occur. Relative motions can also interfere with the astronomical observations and their interpretation. In fact, when looking back to the history of astronomy, one finds a number of incidents where relative-motion effects have led us to a gross misinterpretation of reality. Here, we will look into  the possibility that analogous kinematic illusions can take place on cosmological scales as well.

We are moving observers in the universe. Not only within our solar-system and our galaxy, but on cosmological scales too. The Milky Way and the Local Group, for example, are moving relative to the rest-frame of the cosmos, which is believed to coincide with the coordinate system of the Cosmic Microwave Background (CMB) photons, at approximately 600~km/sec (e.g.~see~\cite{Ketal}). Moreover, surveys of cosmic velocity fields have repeatedly reported bulk peculiar flows with velocities ranging between few hundred and several hundred km/sec and with sizes extending from few hundred to several hundred Mpc (e.g.~see~\cite{Aetal} and references therein). Although several of the reported bulk flows are consistent with the current cosmological model, a number of them are too large and too fast to comply with the $\Lambda$CDM paradigm~\cite{KA-BKE}-\cite{SYFB}. Among the latter is a recent report utilising data from the \textit{CosmicFlows-4} catalogue, claiming a bulk flow of approximately 250~Mpc and velocity in excess of 400~km/sec~\cite{Wetal}. In summary, it seems fair to say that no real observer in the universe follows the idealised CMB frame, but that we all have some finite peculiar velocity relative to it.

Peculiar motions are matter in motion and moving matter implies nonzero energy flux. As a result, in the presence of peculiar velocities, the cosmic medium cannot be treated as perfect. The ``imperfection'' will appear in the form of a nonzero energy-flux vector, which is solely triggered by the peculiar motion of the matter and survives at the linear perturbative level~\cite{TCM,EMM}. In general relativity, as opposed to Newtonian physics, energy fluxes gravitate, since they also contribute to the energy-momentum tensor.  In a sense, one could argue that peculiar flows ``gravitate''~\cite{TT}. It is the inevitable emergence of a nonzero energy flux and its subsequent gravitational input that makes the difference between Newtonian theory and general relativity, when it comes to the study of large-scale peculiar flows and their implications.

The role of the \textit{peculiar flux} becomes clear when we use the so-called ``tilted'' cosmological models. These allow for two (at least) families of relatively moving observers and are therefore best suited to study peculiar motions. Typically, one group are the idealised observers following the reference CMB frame, while the other are their real counterparts moving along with the matter. Then, even when the peculiar velocities are non-relativistic, the relative motion inevitably leads to a nonzero energy flux. This adds to the (perturbed) stress-energy tensor and, in so doing, it modifies the linear conservation laws and eventually leads to flux-related terms in the equations monitoring the evolution and the effects of the peculiar-velocity field.

Peculiar velocities have been largely bypassed in cosmological studies and, when included, the analysis is typically Newtonian. So far, the fully relativistic treatments have led to faster than expected linear growth rates for the peculiar-velocity field~\cite{TT}-\cite{MiT}. These results support recent surveys (e.g.~see~\cite{Wetal}) reporting faster and deeper bulk peculiar flows than those allowed by the current cosmological model. With the exception of the dipole, peculiar velocities are not expected to affect the CMB spectrum, given their late (post-recombination) origin. On the other hand, in principle, bulk peculiar motions could have played a role during structure formation and they could also induce dipole-like anisotropies in the number counts of distant astrophysical sources, like those reported in~\cite{Setal,S} for example. This remain to be seen, however, since there are no specific theoretical studies as yet.

Among the theoretical formulae affected by the aforementioned relativistic gravitational input of the peculiar flux is the Raychaudhuri equation, which determines the deceleration/acceleration rate of the universe. The implications of the observer's peculiar motion for the deceleration parameter measured in their own rest-frame, were originally discussed in~\cite{T1,T2} and later refined in~\cite{TK,T3}, using linear relativistic perturbation theory on an Einstein-de Sitter background. In all cases, a linear velocity tilt was introduced to account for the observer's peculiar flow. It was shown there that observers living inside large-scale bulk flows, like those reported in~\cite{WFH}-\cite{Wetal} for example, can assign a negative value to their locally measured deceleration parameter, while the surrounding universe is globally decelerating. Put another way, in the \textit{tilted-universe scenario}, the recent cosmic acceleration may be a mere illusion triggered by the observer's peculiar motion relative to the reference (CMB) frame of the universe. This can happen when the bulk-flow domain is expanding slightly slower than its surroundings, in which case the unsuspecting observers could easily misinterpret their slower local expansion-rate as recent acceleration of the host universe.   Subsequent analysis, confirmed the above result, this time on general Friedmann-Robertson-Walker (FRW) backgrounds with nonzero pressure and any type of spatial curvature~\cite{T4}. Although, the pressure and the curvature introduced new ingredients, these remained subdominant. In all realistic scenarios, the key feature of the relative-motion effect, namely its scale-dependence, remained the same and dominated on sub-horizon lengths. More specifically, the \textit{transition length}, namely the scale/redshift where the locally measured deceleration parameter appears to turn negative was found to be independent of the Friedmann background. Here, we extend the study to perturbed Bianchi-type universes. Our aim is to test whether the standard (FRW) picture of the bulk-flow effects, as described in~\cite{T1}-\cite{T4} and outlined above, persists despite the spatial anisotropy of the Bianchi spacetimes, or not.

The Bianchi models are the natural anisotropic extensions of the Friedmann universes and, as such, they have provided the testing ground for the cosmological implications of spatial anisotropy. In fact, recent reports of dipolar anisotropies in the number counts of distant astrophysical sources (e.g.~see~\cite{Setal,S}), have drawn renewed attention to the Bianchi cosmological models and to the potential analogies their intrinsically anisotropic nature may have with our cosmos~\cite{KMS-J}. Typically, the Bianchi spacetimes are classified in two types, namely in the non-tilted and the tilted models~\cite{WE}. In the former class, the worldlines of the matter are orthogonal to the hypersurfaces of homogeneity, while in the latter this is no longer the case. Also note that several of the Bianchi universes have the Friedmann models as special cases (see \S~\ref{sBUs} below for more details and references).

In this work, we will perturb a non-tilted Banchi background, so that the tilt is a linear velocity perturbation reflecting the peculiar motion of the real observers relative to the CMB frame. Assuming zero pressure, we calculate the effects of bulk peculiar flows on the deceleration parameter, as measured by the associated observers. Not surprisingly, we find that the background anisotropy adds extra degrees of freedom to the original scenario. The shear input to the relative-motion effect modifies the value of the deceleration parameter measured by the bulk-flow observers. In fact, when the anisotropy is high, the shear contribution can even change the sign of the deceleration parameter from positive to negative. Nevertheless, for realistic background anisotropy, the aforementioned shear-related effects are subdominant and the Einstein-de Sitter scenario reappears. Observers in perturbed Bianchi universes residing in bulk peculiar flows, like those reported in~\cite{WFH}-\cite{Wetal} can still have the illusion of accelerated  expansion in a decelerating universe. Just like in the Einstein-de Sitter and the generalised FRW studies, the key feature of the peculiar-motion effect, namely its scale-dependence, remains the same and dominates on sub-horizon lengths. The domain of dominance is determined by the associated transition length, which ranges between few hundred and several hundred Mpc~\cite{T3,T4}. The latter is large enough to make a local relative-motion effect look like a recent global event and, in so doing, mislead the unsuspecting bulk-flow observers to the illusion of universal acceleration.

\section{Bianchi universes}\label{sBUs}
The (non-tilted) Bianchi universes are a family of ten spatially homogeneous cosmological models that allow for spatial anisotropy. Five of these Bianchi spacetimes also contain the Friedmann universes, with all three types of spatial curvature, as special cases (see Table~\ref{tab1}).

\begin{table}
\caption{The non-tilted Bianchi models classified into two group classes and ten group types~\cite{WE}.}
\vspace{0.5truecm}
\begin{center}\begin{tabular}{cccccc}
\hline \hline \hspace{-2.5mm}  Group class \hspace{-2.5mm} & Group type \hspace{-2.5mm} & $n_1$ & $n_2$ & $n_3$ & \hspace{-2.5mm} FRW special case \hspace{-2.5mm} \\ \hline A$\,(a=0)$ & $\left\{\begin{array}{l} I \\ II \\ VI_0 \\ VII_0 \\ VIII \\ IX \end{array}\right.$ & $\begin{array}{c} 0 \\ + \\ 0 \\ 0 \\ - \\ + \end{array}$ & $\begin{array}{c} 0 \\ 0 \\ + \\ + \\ + \\ + \end{array}$ & $\begin{array}{c} 0 \\ 0 \\ - \\ + \\ + \\ + \end{array}$ & $\begin{array}{c} K=0 \\ - \\ - \\ K=0 \\ - \\ K=+1 \end{array}$ \\[1.5truemm]\hline B$\,(a\neq0)$ & $\left\{\begin{array}{l} V \\ IV \\ VI_h \\ VII_h \end{array}\right.$ & $\begin{array}{c} 0 \\ 0 \\ 0 \\ 0 \end{array}$ & $\begin{array}{c} 0 \\ 0 \\ + \\ + \end{array}$ & $\begin{array}{c} 0 \\ + \\ - \\ + \end{array}$ & $\begin{array}{c} K=-1 \\ - \\ - \\ K=-1 \end{array}$ \\ [1.5truemm] \hline \hline
\end{tabular}\end{center}\label{tab1}\vspace{0.5truecm}
\end{table}

\subsection{The background equations}\label{ssBEs}
The Bianchi solutions go beyond the familiar (spatially homogeneous and isotropic) Friedmann models, since they can naturally support kinematic and dynamic anisotropies. The geometry of the Bianchi spacetimes is also generically anisotropic, allowing for a nonzero Weyl field and for anisotropic 3-Ricci curvature. As a result, Bianchi-type cosmologies have been used to investigate the role and the implications of spatial anisotropy for the evolution of the universe.\footnote{Recent reports of dipolar anisotropies in the number counts of distant astrophysical sources, like quasars for example~\cite{Setal}, have drawn renewed attention to the Bianchi cosmological models and to the potential analogies their intrinsically anisotropic nature may have with our cosmos~\cite{KMS-J}.}

Focusing on those Bianchi universes that contain the FRW solutions as special cases, let us set the cosmological constant to zero and assume matter in the perfect fluid form (with density $\rho$ and isotropic pressure $p$). Then, introducing a group of (fundamental) observers with timelike 4-velocity $u_a$ (so that $u_au^a=-1$), the volume scalar $\Theta=\nabla^au_a$ defines the observers' mean expansion rate and the associated Hubble parameter ($H$, with $H=\Theta/3$ by construction). The latter in turn determines the corresponding scale factor $a=a(t)$ by means of $\dot{a}/a=H$. The curvature of the spatial hypersurfaces orthogonal to the $u_a$-field is determined by the 3-curvature index $K=0,\pm1$. Also, $\sigma^2= \sigma_{ab}\sigma^{ab}/2$ is the magnitude of the symmetric and trace-free shear tensor ($\sigma_{ab}$), which measures the kinematic anisotropy of the models. Employing the above, the generalised Friedmann and Raychaudhuri formulae read
\begin{equation}
H^2= {1\over3}\left(\rho+\sigma^2\right)- {K\over a^2} \hspace{15mm} {\rm and} \hspace{15mm} \dot{H}= -H^2- {1\over6}\left(\rho+3p\right)- {2\over3}\,\sigma^2\,,  \label{Bianchi}
\end{equation}
respectively. An additional key relation is the continuity equation, which in a Bianchi universe filled with an ideal medium takes the form
\begin{equation}
\dot{\rho}= -3H(\rho+p)\,.  \label{Bcont}
\end{equation}
Note that, on introducing the density parameters $\Omega_{\rho}=\rho/3H^2$, $\Omega_K=-K/(aH)^2$ and $\Omega_{\sigma}=\sigma^2/3H^2$, the generalised Friedmann and Raychaudhuri equations respectively recast as
\begin{equation}
1= \Omega_{\rho}+ \Omega_K+ \Omega_{\sigma} \hspace{15mm} {\rm and} \hspace{15mm} q= {1\over2}\,(1+3w)\Omega_{\rho}+ 2\Omega_{\sigma}\,,  \label{Omegas}
\end{equation}
where $q=-[1+(\dot{H}/H^2)]$ is the deceleration parameter and $w=p/\rho$ is the barotropic index of the matter. Note that, in both of the above, the dimensionless parameter $\Omega_{\sigma}$ provides a measure of the anisotropy of the Bianchi model.

\subsection{Perturbing the Bianchi background}\label{ssPBB}
One can study the role of spatial anisotropy during the linear stages of structure formation by perturbing a given Bianchi background. This means allowing for weak inhomogeneity in all the terms of the linear equations. For instance, at the linear perturbative level, the Friedmann and the Raychaudhuri formulae read~\cite{TCM,EMM}
\begin{equation}
\mathcal{R}= 2\left(\rho-{1\over3}\,\Theta^2+\sigma^2\right)  \label{lBianchi1}
\end{equation}
and
\begin{equation}
\dot{\Theta}= -{1\over3}\,\Theta^2- {1\over2}\,(\rho+3p)- 2\sigma^2+ {\rm D}^aA_a\,,  \label{lBianchi2}
\end{equation}
respectively. In the above, $\mathcal{R}$ is the perturbed spatial Ricci scalar and $A_a=\dot{u}_a$ is the linear 4-acceleration vector, with ${\rm D}_a=h_a{}^b\nabla_b$ being the 3-dimensional covariant derivative operator. Also, $h_{ab}=g_{ab}+u_au_b$ is the symmetric projection tensor (with $g_{ab}$ being the spacetime metric and $h_{ab}u^b=0$), which also acts as the metric of the 3-dimensional spaces orthogonal to the $u_a$-field.

The linear Raychaudhuri equation is key to our analysis because it determines the deceleration/acceleration rate of the universe. Indeed, recalling that $q=-([1+(3\dot{\Theta}/\Theta^2)]$ by definition, expression (\ref{lBianchi2}) recasts into
\begin{equation}
{1\over3}\,\Theta^2q={1\over2}\,(\rho+3p)+ 2\sigma^2- {\rm D}^aA_a\,.  \label{lBq}
\end{equation}
Hence, for conventional matter (with $\rho+3p>0$), the expansion is always decelerated, unless the 4-acceleration term seen on the right-hand side of the above can somehow reverse the situation. The 4-acceleration is given by the momentum-density conservation law, which on a Bianchi background linearises to~\cite{TCM,EMM}
\begin{equation}
(\rho+p)A_a= -{\rm D}_ap- \dot{q}_a- 4Hq_a-\sigma_{ab}q^b- {\rm D}^a\pi_{ab}\,,  \label{lBmcl}
\end{equation}
with $q_a$ and $\pi_{ab}$ representing the linear energy flux and viscosity of the matter respectively. Following (\ref{lBmcl}), the 4-acceleration is zero in perturbed (non-tilted) Bianchi universes filled with an ideal pressureless fluid. However, in the presence of peculiar motions, there is always a nonzero energy flux due to the motion of the matter, which leads to a nonzero 4-acceleration (see \S~\ref{ssPFP4-A} below). Put another way, in linearised tilted cosmologies (Bianchi and Friedmann) the 4-acceleration is not zero, even in the absence of pressure. In what follows, we will demonstrate the key role of the non-vanishing 4-acceleration in perturbed Bianchi universes with a peculiar-velocity tilt.

\section{Perturbed Bianchi universes with a tilt}\label{sPBUT}
Let us consider a perturbed Bianchi spacetime and also allow for weak peculiar velocity perturbations. The latter induce a nonzero tilt to the perturbed spacetime, which can be treated as an additional dynamical variable.

\subsection{Relatively moving observers}\label{ssRMOs}
\begin{figure}[!tbp]
\centering\includegraphics[width=0.5\textwidth]{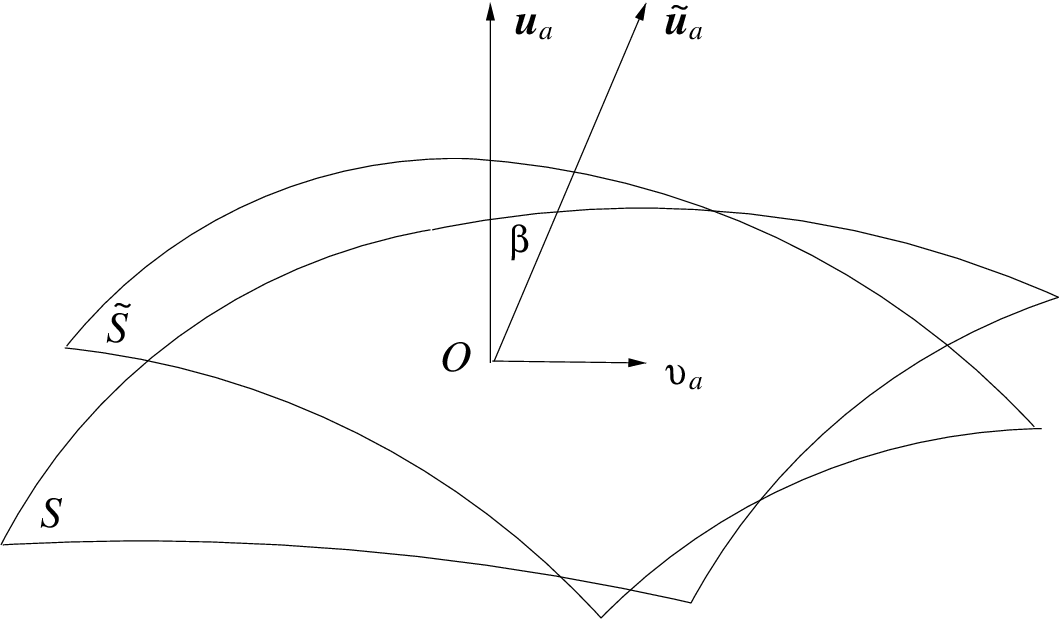}
  \caption{Tilted cosmologies have two groups of relatively moving observers, with 4-velocities $u_a$ and $\tilde{u}_a$, at every spacetime event ($O$). Identifying the $u_a$-field with the reference (CMB) coordinate system of the universe, makes $\tilde{u}_a$ the 4-velocity of the real observers, moving with peculiar velocity $v_a$ relative to the CMB. Also, $\beta$ (with $\cosh\beta=-u_a\tilde{u}^a$) is the hyperbolic (tilt) angle between $u_a$ and $\tilde{u}_a$, while $S$ and $\tilde{S}$ are the 3-D rest-spaces of the two observer groups.}  \label{fig:pmotion}
\end{figure}

In cosmology, we define and measure peculiar velocities with respect to the reference system of the universe, which has been traditionally identified with the rest-frame of the cosmic microwave photons. By default, the latter is the only coordinate system where the radiation dipole vanishes (e.g.~see~\cite{EMM}). However, only idealised observers follow the CMB frame. Real observers, living in typical galaxies like our Milky Way, follow a ``tilted'' system that moves relative to the microwave background with finite peculiar velocity (see Fig.~\ref{fig:pmotion}).\footnote{The CMB frame is quite often referred to as the Hubble frame, though strictly speaking they are distinct.} Due to their peculiar motion, the real observers see an (apparent) dipolar anisotropy in the CMB spectrum, when their idealised counterparts should see none. On these grounds and assuming that the observed CMB dipole is purely kinematic, the peculiar velocity of our Local Group is close to 600~km/sec.

When the peculiar motions are non-relativistic, the 4-velocity of the real observers ($\tilde{u}_a$) and that of their idealised counterparts ($u_a$) are related by
\begin{equation}
\tilde{u}_a= u_a+ v_a\,,  \label{4vels}
\end{equation}
where $v_a$ is the peculiar velocity of the former observers relative to the latter (e.g.~see~\cite{TCM,EMM} and Fig.~\ref{fig:pmotion} here). Note that $u_av^a=0$ by construction and $v^2=v_av^a\ll1$ at the non-relativistic limit. Also, both coordinate systems live in the perturbed universe and we may use either of them to analyse the peculiar-velocity field and its implications. Here, we will take the perspective of the CMB observers.\footnote{Taking the viewpoint of the real (the tilted) observers makes no difference to the final results (see~\cite{T3,T4}), since all that matters is the relative motion between the two frames.}

The nature of the cosmic medium experienced by the relatively moving observers differs, even at the linear level. Indeed, assuming a Bianchi background filled with a single perfect fluid, the matter variables measured in the two coordinate systems are related by~\cite{M}
\begin{equation}
\tilde{\rho}= \rho\,, \hspace{10mm} \tilde{p}= p\,, \hspace{10mm} \tilde{\pi}_{ab}= \pi_{ab}  \label{mlrels1}
\end{equation}
and
\begin{equation}
\tilde{q}_a= q_a- (\rho+p)v_a\,,  \label{mlrels2}
\end{equation}
to first approximation. In the above, $\rho$, $p$, $\pi_{ab}$ and $q_a$ are respectively the density, the pressure, the viscosity and the energy flux of the matter measured in the reference (CMB) coordinate system, with their tilted counterparts measured in the frame of the matter. Similarly, the linear kinematic variables measured in two frames are related by
\begin{equation}
\tilde{\Theta}= \Theta+ \vartheta\,, \hspace{10mm} \tilde{\sigma}_{ab}= \sigma_{ab}+ 2u_{(a}\sigma_{b)c}v^c+ \varsigma_{ab}\,,   \label{klrels1}
\end{equation}
\begin{equation}
\tilde{\omega}_{ab}= \omega_{ab}+ \varpi_{ab}  \label{klrels2}
\end{equation}
and
\begin{equation}
\tilde{A}_a= A_a+ \dot{v}_a+ Hv_a+ \sigma_{ab}v^b\,.  \label{klrels3}
\end{equation}
Here, $\Theta$, $\sigma_{ab}$, $\omega_{ab}$ and $A_a$ are respectively, the expansion scalar, the shear tensor, the vorticity tensor and the 4-acceleration vector measured in the CMB frame, while their tilted counterparts are evaluated in that of the matter. Also, $\vartheta={\rm D}^av_a$, $\varsigma_{ab}={\rm D}_{\langle b}v_{a\rangle}$ and $\varpi_{ab}={\rm D}_{[b}v_{a]}$ are respectively the expansion/contraction scalar, the shear tensor and the vorticity tensor of the peculiar motion.\footnote{Angled brackets denote symmetric and traceless tensors. For example, ${\rm D}_{\langle b}v_{a\rangle}={\rm D}_{(b}v_{a)}-(\vartheta/3)h_{ab}$.} Note that, whereas both $\Theta$ and $\tilde{\Theta}$ are positive, $\vartheta$ is positive in a locally expanding bulk flow and negative when the latter is contracting. It goes without saying that, at the linear perturbative level, $|\vartheta|\ll\Theta$ by default.

\subsection{Peculiar flux and peculiar 
4-acceleration}\label{ssPFP4-A}
Of particular interest for the purposes of this study, are relations (\ref{mlrels2}) and (\ref{klrels3}). According to these linear formulae, peculiar motions imply nonzero energy flux and nonzero 4-acceleration. Indeed, even when these variables vanish in one frame, they will both take nonzero values in every other relatively moving coordinate system. The frame choice makes no physical difference, since all that matters is their relative motion (see footnote~2 here and also~\cite{T3,T4}). Assuming a pressureless cosmic medium, we will take the ``traditional'' approach and set $\tilde{q}_a=0$ and $\tilde{A}_a=0$ in the matter frame~\cite{M}-\cite{ET}. In other words, there is no flux in the frame of the matter, which moves along timelike geodesics. Then, expressions (\ref{mlrels2}) and (\ref{klrels3}) recast to
\begin{equation}
q_a= \rho v_a \hspace{7.5mm} {\rm and} \hspace{7.5mm} A_a= -\dot{v}_a- Hv_a- \sigma_{ab}v^b\,,  \label{lrels3}
\end{equation}
in the reference coordinate system. Comparing to the FRW studies, the only difference is the extra shear term on the right-hand side of (\ref{lrels3}b). Before we proceed to investigate the implications of the above relations, it is worth taking a closer look into the physics behind them.

Following (\ref{mlrels2}) and (\ref{lrels3}a), peculiar velocities ensure a nonzero \textit{peculiar flux}. The only exception is when the cosmic medium has an de Sitter inflationary equation of state, with $p=-\rho$ (see~\cite{MaT} for further discussion). Nonzero peculiar flux makes a pivotal difference, because the cosmic fluid can no longer be treated as perfect (e.g.~see \S~5.2.1 in~\cite{EMM}). This in turn implies nonzero 4-acceleration even when there is no pressure (see~\cite{TCM,EMM} and also below). It all happens because the peculiar flux contributes to the (perturbed) energy-momentum tensor and therefore to the gravitational field. The flux input feeds into the relativistic conservations laws of energy and momentum (see Eqs.~(1.3.17) and (1.3.18) in~\cite{TCM}, or (5.11) and (5.12) in~\cite{EMM} for the nonlinear expressions), which in a tilted Bianchi universe with pressureless matter linearise to
\begin{equation}
\dot{\rho}= -\Theta\rho- {\rm D}^aq_a  \label{lBcls1}
\end{equation}
and
\begin{equation}
\rho A_a= -\dot{q}_a- 4Hq_a- \sigma_{ab}q^b\,,  \label{lBcls2}
\end{equation}
respectively. The momentum-conservation law (\ref{lBcls2}) seen above reveals the direct interconnection between the \textit{peculiar flux} and the \textit{peculiar 4-acceleration}.\footnote{Substituting (\ref{lrels3}a) into the momentum density conservation law (\ref{lBcls2}) recovers expression (\ref{lrels3}b).} Crucially, this holds in the absence of pressure. Then, the form of the 4-acceleration, which in relativity describes non-gravitational forces, follows naturally from standard linear cosmological perturbation theory. Indeed, using (\ref{lrels3}a) and the linear commutation law ${\rm D}_a\dot{\rho}=({\rm D}_a\rho)^{\cdot}+H{\rm D}_a\rho+3H\rho A_a+\sigma_{ab}{\rm D}^b\rho$, the spatial gradient of (\ref{lBcls1}) gives
\begin{equation}
A_a= -{1\over3H}\,{\rm D}_a\vartheta- {1\over3aH}\left(\dot{\Delta}_a+\mathcal{Z}_a+ \sigma_{ab}\Delta^b\right)\,,  \label{lA}
\end{equation}
where $\Delta_a=(a/\rho){\rm D}_a\rho$ and $\mathcal{Z}_a=a{\rm D}_a\Theta$ monitor inhomogeneities in the matter and in the expansion respectively (e.g.~see~\cite{TCM,EMM}). The above is the fully relativistic linear expression for the peculiar 4-acceleration in a tilted, almost-Bianchi universe that is filled with pressureless matter and in the presence of peculiar velocities.\footnote{An alternative way of obtaining (\ref{lA}) is by linearising the nonlinear expression (2.3.1) of~\cite{TCM}, or equivalently Eq.~(10.101) of~\cite{EMM}, on our Bianchi background and then using (\ref{lBcls2}) to  substitute for the 4-acceleration.} Thus, the momentum conservation law (\ref{lBcls2}) guarantees the presence of a nonzero peculiar 4-acceleration and the energy conservation law (\ref{lBcls1}) provides the relativistic expression for the peculiar 4-acceleration. In what follows, we will consider the implications of (\ref{lA}) for the kinematics of the aforementioned Bianchi-type universe.

Before doing so, we should emphasise that everything stems from the contribution of the peculiar flux to the gravitational field. The effect is purely general relativistic in nature, it ``survives'' at the linear perturbative level and it cannot be reproduced in Newtonian, as well as in quasi-Newtonian, studies.\footnote{Although the quasi-Newtonian studies of peculiar velocities have an apparently relativistic profile, they reduce to Newtonian for all practical purposes. The reason is the adopted reference frame, which has zero linear shear and vorticity (e.g.~\cite{M}-\cite{ET}). This imposes severe constraints on the perturbed spacetime (e.g.~no gravitational waves either) and eventually leads to Newtonian-like equations and results (see \S~6.8.2 in~\cite{EMM} for ``warning'' comments). The ``apparent'' advantage is that, in the absence of shear and vorticity, one is allowed to express the linear 4-acceleration as the gradient $A_a={\rm D}_a\varphi$ of a scaler potential $\varphi$~\cite{M}-\cite{ET}. The latter, however, is arbitrary and essentially identical to its purely Newtonian counterpart. As a result, in the quasi-Newtonian studies the gravitational input of the peculiar flux is never accounted for, which means that the fully relativistic relation (\ref{lA}) for the 4-acceleration is entirely bypassed. For further discussion and for a direct comparison between the relativistic and the Newtonian/quasi-Newtonian studies of peculiar motions the reader is referred to~\cite{T5}.} Even proper relativistic treatments of cosmological peculiar motions can (inadvertently) miss the gravitational input of the peculiar flux, when they employ the so-called energy (or Landau-Lifshitz, or centre-of-mass) frame, where the flux vector vanishes by default (e.g.~see \S~3.3.3 and \S~3.3.4 in~\cite{TCM}, or \S~10.4.3 i~\cite{EMM}, as well as~\cite{CMV-R}). Overall, accounting for peculiar-flux contribution to the gravitational field is what distinguishes the relativistic studies of peculiar velocities from the rest (see~\cite{T5} for further discussion).

\section{Relative-motion effects on the deceleration 
parameter}\label{sRMEDP}
The presence of a peculiar 4-acceleration in one of the two relatively moving frames, means that the Raychaudhuri equations in these coordinate systems differ. Then, the deceleration parameters measured by the idealised (CMB) and the real (tilted) observers will differ as well.

\subsection{Two deceleration parameters}\label{ssTDPs}
In the absence of pressure, the Raychaudhuri equations (see expression (\ref{lBq}) in \S~\ref{ssPBB}), in CMB and in the matter frame of our tilted almost-Bianchi universe, linearises to
\begin{equation}
{1\over3}\,\Theta^2q= {1\over2}\,\rho+ 2\sigma^2- {\rm D}^aA_a   \label{lBCMBRay}
\end{equation}
and
\begin{equation}
{1\over3}\,\tilde{\Theta}^2\tilde{q}= {1\over2}\,\tilde{\rho}+ 2\tilde{\sigma}^2\,,  \label{lBTRay}
\end{equation}
respectively (recall that $A_a\neq0$ and $\tilde{A}_a=0$ -- see \S~\ref{ssPFP4-A} earlier). In the above, $q$ is the deceleration parameter measured in the CMB frame and $\tilde{q}$ is the one measured by the real observers in a typical galaxy like our Milky Way. The difference between Raychaudhuri equations, as reflected in (\ref{lBCMBRay}) and (\ref{lBTRay}), suggests that the associated deceleration parameters should differ too. Indeed, since $\tilde{\Theta}=\Theta+\vartheta$, $\tilde{\rho}=\rho$, $\tilde{\sigma}^2=\sigma^2+ \sigma_{ab}\varsigma^{ab}$ and $|\vartheta|/\Theta\ll1$ to first approximation, the aforementioned formulae combine to provide the following linear relation
\begin{equation}
\tilde{q}- q= {2\over3H^2}\,\sigma_{ab}\tilde{\varsigma}^{ab}+ {1\over3H^2}\,{\rm D}^aA_a\,,  \label{ltq1}
\end{equation}
between the two deceleration parameters. Hence, $\tilde{q}\neq q$ and the difference is entirely due to relative-motion effects. The latter also determine the magnitude of the ``correction term'' seen on the right-hand side of the above. Note that, compared to the FRW cases discussed in~\cite{TK}-\cite{T4}, the anisotropy of the Bianchi background has added a shear contribution to the correction term.

\subsection{The correction term}\label{ssCT}
Let us start by focusing on the contribution of the 4-acceleration  to the difference between the two deceleration parameters, as seen in Eq.~(\ref{ltq1}) above. Taking the spatial divergence of (\ref{lA}) and using the linear commutation law $a{\rm D}^a\dot{\Delta}_a=\dot{\Delta}+\sigma_{ab}\Sigma^{ab}$, we obtain
\begin{equation}
{\rm D}^aA_a= -{1\over3H}\,{\rm D}^2\vartheta- {1\over3a^2H}\left(\dot{\Delta}+\mathcal{Z} +2\sigma_{ab}\Sigma^{ab}\right)\,.  \label{ldivA}
\end{equation}
Here, ${\rm D}^2={\rm D}^a{\rm D}_a$ is the covariant spatial Laplacian operator, while $\Delta= a{\rm D}^a\Delta_a$, $\mathcal{Z}=a{\rm D}^a\mathcal{Z}_a$ and $\Sigma_{ab}=a{\rm D}_{\langle b}\Delta_{b\rangle}$ by definition~\cite{TCM,EMM}. Substituting the above into the right-hand side of (\ref{ltq1}), the latter reads
\begin{equation}
\tilde{q}- q= {2\over3H^2}\,\sigma_{ab}\tilde{\varsigma}^{ab}- {1\over9H^3}\,{\rm D}^2\vartheta- {1\over9H}\left({\lambda_H\over\lambda_K}\right)^2 \left(\dot{\Delta}+\mathcal{Z}+2\sigma_{ab}\Sigma^{ab}\right)\,,  \label{ltq2}
\end{equation}
with $\lambda_H=1/H$ representing the Hubble radius and $\lambda_K=a$ the curvature scale of the Bianchi background -- e.g.~see Table~\ref{tab1}. Given that $\lambda_H/\lambda_K\ll1$ according to the observations, we may ignore the last term on the right-hand side of the above.\footnote{Even if one was to keep the last term on the right-hand side of (\ref{ltq2}), the results will remain unchanged for all practical purposes (see~\cite{T4} for further discussion and details).} Also, the spatial Laplacian suggests a scale dependence, which becomes explicit after a harmonic decomposition. Then, (\ref{ltq2}) reduces to
\begin{equation}
\tilde{q}- q= {2\over3H^2}\,{\rm D}^a\left(\sigma_{ab}v^b\right)+ {1\over9}\left({\lambda_H\over\lambda}\right)^2{\vartheta\over H}\,,  \label{ltq3}
\end{equation}
where $\lambda=\lambda_n=a/n$ is the scale of the bulk peculiar flow and we have dropped the harmonic index ($n$) from Eq.~(\ref{ltq3}) for the economy of the presentation. On the right-hand side of the above, we have the correction term solely induced by the peculiar motion of the matter frame relative to the CMB. Expression (\ref{ltq3}) reproduces the linear FRW-formula of~\cite{TK}-\cite{T4}, with an additional shear term due to our Bianchi background.

\subsection{The overall effect on $\tilde{q}$}\label{ssOEtq}
Without compromising the impact of the anisotropy, we may assume that the peculiar velocity is a shear eigenvector, so that $\sigma_{ab}v^b= \zeta v_a$, with $\zeta$ being the associated eigenvalue.\footnote{The expansion rate of an anisotropic Bianchi universe varies between different directions and the difference is monitored by the expansion tensor, defined by $\Theta_{ab}=(\Theta/3)h_{ab}+\sigma_{ab}$ (e.g.~see~\cite{TCM,EMM}). Then,  $\Theta_{ab}v^b= Hv_a+\sigma_{ab}v^b$ to linear order. When $v_a$ is a shear eigenvector, with $\sigma_{ab}v^b= \zeta v_a$, the last relation becomes $\Theta_{ab}v^b=(H+\zeta)v_a$. Accordingly, when $\zeta>0$, the shear input enhances the background expansion in the direction of the bulk velocity ($v_a$), whereas in the opposite case the expansion along that direction slows down. Note that, in Bianchi cosmologies with a observationally realistic degree of anisotropy, we have $|\zeta|/H<1$ always.} Then, recalling the ${\rm D}^av_a=\vartheta$, expression (\ref{ltq3}) recasts into
\begin{equation}
\tilde{q}= q+ {2\over3}\left[{\zeta\over H} +{1\over6}\left(\lambda_H\over\lambda\right)^2\right] {\vartheta\over H}\,.  \label{ltq4}
\end{equation}
This is the deceleration parameter measured by the tilted observers, living inside bulk flows in a perturbed, nearly-flat Bianchi universe with dust and no cosmological constant. Compared to the FRW studies (e.g.~see Eqs.~(13), (21) in~\cite{T3}, \cite{T4} respectively), the background anisotropy has only added the linear $\zeta/H$ term. Therefore, the overall effect of the observer's peculiar flow on $\tilde{q}$ depends on the background shear anisotropy (measured by the $\zeta/H$ ratio) and on the scale of the bulk peculiar flow (determined by $\lambda_H/\lambda$). Following (\ref{ltq4}), the relative-motion effect is more sensitive to the scale-ratio $(\lambda_H/\lambda)$ than to the background anisotropy. The last, but not the least, decisive factor is the sign of the peculiar volume scalar ($\vartheta$). Recall that positive values of $\vartheta$ denote locally expanding bulk flows, while the negative ones imply local contraction.

According to Eq.~(\ref{ltq4}), the correction term due to the relative-motion increases when $\zeta>0$, while it decreases for $\zeta<0$. This means that the relative-motion effect on $\tilde{q}$ is stronger when the shear adds to the background expansion-rate (see footnote~8). Also, as in the FRW cases discussed in~\cite{TK}-\cite{T4}, the scale-ratio ($\lambda_H/\lambda$) ensures that bulk-flow effect becomes stronger on progressive smaller lengths and fades away (as expected) at high redshifts. In principle, one may consider a broad range of scenarios, with different combinations of $\zeta/H$ and $\lambda_H/\lambda$, to discuss the impact of the observer's peculiar flow. Next, we will look at the two most characteristic alternatives.\\

\textbf{(i)} On sufficiently large scales the impact of peculiar motions is expected to fade away, as the associated velocities tend to zero on sufficiently long wavelengths. This is also reflected in Eq.~(\ref{ltq4}), since the scale ratio seen on its right-hand side decreases quickly on progressively larger lengths. For instance, confining to super-horizon lengths, where $\lambda_H/\lambda\ll1$, and assuming that the background anisotropy is high, namely setting $|\zeta|/H\gg1$, expression (\ref{ltq4}) reduces to
\begin{equation}
\tilde{q}= q+ {2\zeta\vartheta\over3H^2}\,.  \label{ltq5}
\end{equation}
Here, the effect on $\tilde{q}$ is solely determined by the sign of $\zeta\vartheta$ and by the overall value of the correction term on the right-hand side of the above. Note that the product $\zeta\vartheta$ carries the combined effect of the background kinematic anisotropy and that of the bulk peculiar flow. Thus, when the shear enhances the background expansion in the direction of a bulk flow that is locally contracting on average, we have $\zeta\vartheta<0$ -- see footnote~8. Then, the value of the local deceleration parameter decreases (i.e.~$\tilde{q}<q$) and it can even drop below zero (in principle). The same also happens when $\vartheta>0$ and $\zeta<0$, namely when the previous kinematic situation is reversed.\\

\begin{figure}[tbp]
\centering \vspace{4cm} \includegraphics{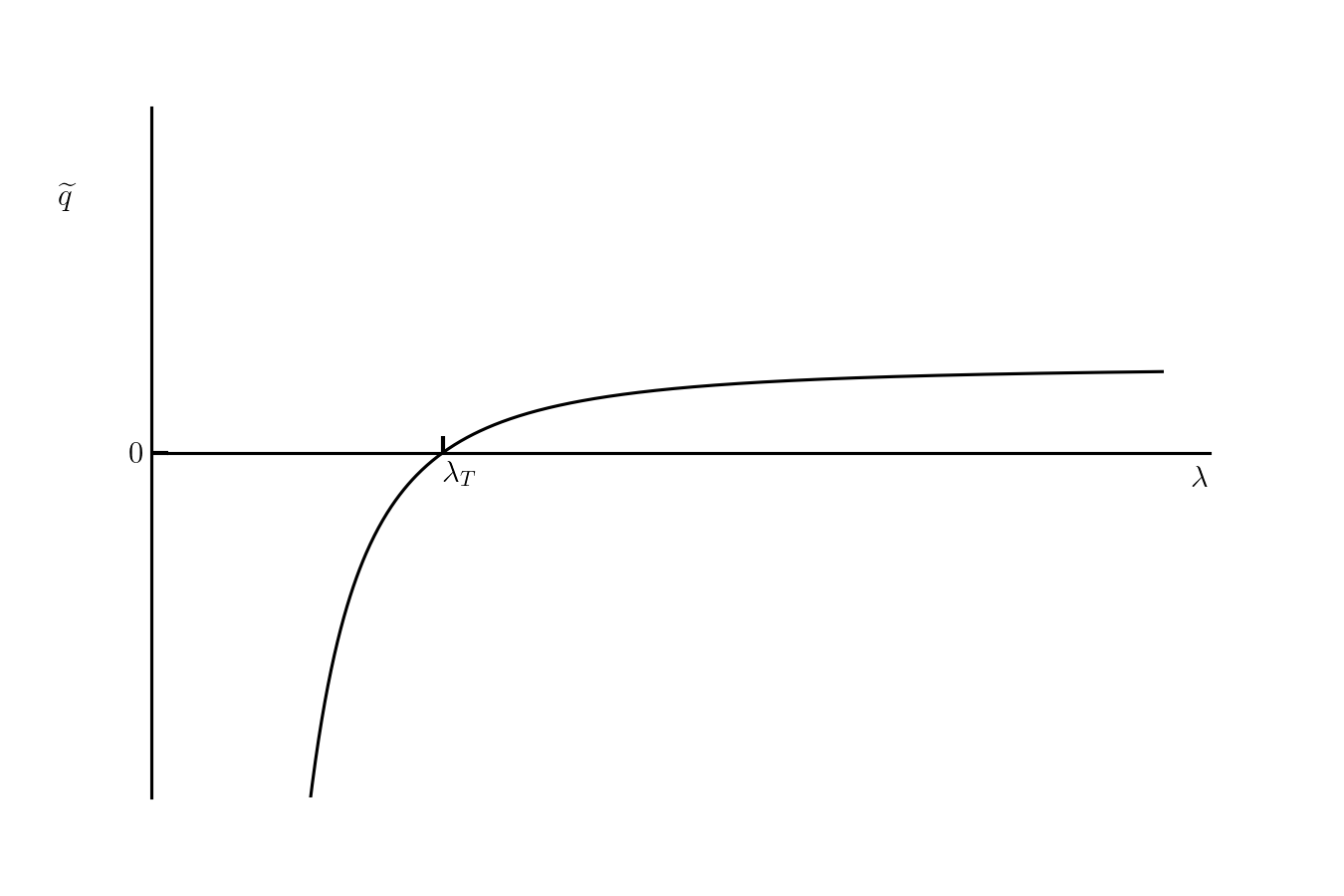} \caption{The linear profile of the deceleration parameter ($\tilde{q}$) measured by observers located at the centre of a locally contracting bulk flow in a tilted almost-Bianchi universe with realistic anisotropy and pressureless dust. The transition scale ($\lambda_T$) marks the threshold where the sign of $\tilde{q}$ appears to change from positive to negative. On scales much larger than the transition length, the locally measured deceleration parameter approaches the CMB value (i.e.~$\tilde{q}\rightarrow q$). The unsuspecting observer may be therefore misled to believe that their universe has recently entered a phase of accelerated expansion. Note that we have assumed subhorizon-sized bulk flows with $\vartheta= \langle\vartheta\rangle=$~constant throughout their interior. Introducing a (physically motivated) scale-dependence to $\vartheta$ does not change the key features of the $\tilde{q}$-profile, but allows to estimate the transition redshift and the current value of $\tilde{q}$ using the SNe~Ia data. For an example of a refined $\tilde{q}$-profile and a comparison to its $\Lambda$CDM counterpart see Fig.~\ref{fig:tq-2} here, as well as Fig.~3 in~\cite{AKPT}.}  \label{fig:tq-1}
\end{figure}

\begin{figure}[tbp]
\centering \vspace{3.5cm} \includegraphics{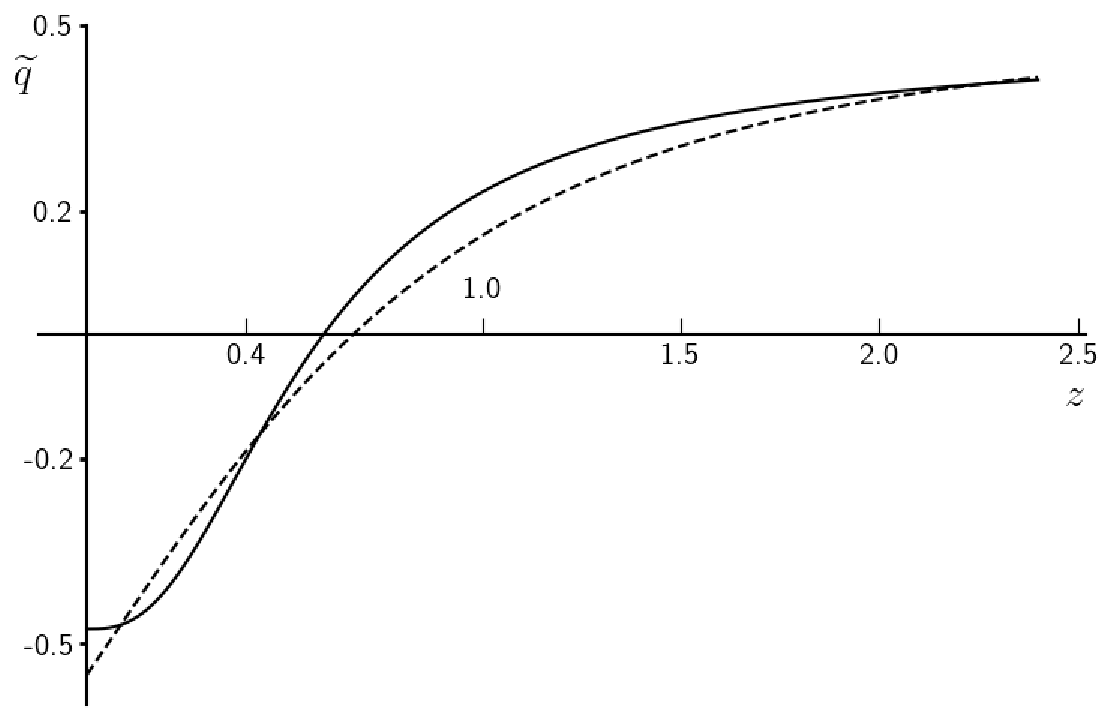} \vspace{20mm}\caption{The linear profile of the deceleration parameter ($\tilde{q}$) in the tilted-universe scenario (solid curve), obtained after allowing for a scale-dependent local volume scalar ($\vartheta= \vartheta(\lambda)$), compared to its $\Lambda$CDM counterpart (dashed curve). The fit was achieved by using the full Pantheon dataset. For further discussion and details the reader is referred to \S~3.2 in~\cite{AKPT}.}  \label{fig:tq-2}
\end{figure}

\textbf{(ii)} Let us now turn to sub-Hubble scales, ranging between few hundred and several hundred Mpc, and consider cosmologically realistic Bianchi backgrounds with weak anisotropy, namely set $\lambda_H/\lambda\gg1$ and $\zeta/H\ll1$. In this scenario, expression (\ref{ltq4}) takes the form
\begin{equation}
\tilde{q}= q+ {1\over9}\left(\lambda_H\over\lambda\right)^2 {\vartheta\over H}\,.  \label{ltq6}
\end{equation}
which coincides with its Einstein-de Sitter and its generalised FRW counterparts (see~\cite{TK,T3} and~\cite{T4} respectively). Here, the effect of the correction term depends crucially on the sign of the peculiar volume scalar $\vartheta$, namely on whether the associated bulk flow is locally expanding or contracting with respect to its surroundings. In the latter case we have $\vartheta<0$, which makes the correction term in (\ref{ltq6}) negative and allows the locally measured deceleration parameter to turn negative as well (i.e.~$\tilde{q}<0$, while $q>0$).\footnote{Assuming that there is not natural bias for contracting or expanding bulk flows on cosmological scales, the chances of an observer residing in either of them should close to 50\%.} Interestingly, a recent reconstruction our local peculiar-velocity field using data from the $2M++$ survey and found that its divergence is  negative~\cite{PGC}. This result, which appears in agreement with the more recent report of~\cite{SDK}, suggests that we may indeed happen to live inside a contracting bulk flow (with $\vartheta<0$) and therefore we could erroneously assign negative values to our local deceleration parameter. In such a case, the \textit{transition length}, namely the scale where the sign of $\tilde{q}$ should appear to change is~\cite{T3}
\begin{equation}
\lambda_T= {1\over3}\sqrt{{|\vartheta|\over qH}} \,\lambda_H\,.  \label{FRWTS}
\end{equation}
Substituting the above into the right-hand side of (\ref{ltq6}), the latter acquires the form~\cite{T3}
\begin{equation}
\tilde{q}= q\left[1-\left(\lambda_T\over\lambda\right)^2\right]\,,  \label{ltq7}
\end{equation}
when the bulk flow is contracting (with $\vartheta<0$). Hence, the deceleration parameter measured by observers in locally contracting bulk flows is positive on scales larger than $\lambda_T$, equals zero at the transition length and turns negative on lengths smaller than $\lambda_T$. Clearly, $\tilde{q}\rightarrow q$ on sufficiently long wavelengths, with $\lambda\gg\lambda_T$ (see also Figs.~\ref{fig:tq-1} and~\ref{fig:tq-2}).\\

Expression (\ref{ltq4}) ensures that, on scales well inside the Hubble radius, the ratio $(\lambda_H/\lambda)^2$ will always dominate the correction term on the right-hand side of (\ref{ltq4}), unless the anisotropy of the Bianchi background is unrealistically high. It is therefore clear that, on sub-horizon lengths with a realistic degree of spatial anisotropy, we always arrive at Eq.~(\ref{ltq6}) and thus recover the results of the tilted Einstein-de Sitter scenario~\cite{TK}-\cite{T4}.

\subsection{The Bianchi transition scale}\label{ssBTS}
Going back to expression (\ref{ltq4}), consider a contracting bulk flow (with $\vartheta<0$) in a perturbed tilted Bianchi universe with non-negligible anisotropy (i.e.~set $|\zeta|/H\lesssim1$). In this more general case, it is possible to make the local deceleration parameter negative, by appealing to both the anisotropy ratio ($\zeta/H$) and to the scale ratio ($\lambda_H/\lambda$). Then, the transition scale, which marks the $\tilde{q}=0$ threshold and the apparent change from deceleration to acceleration, is given by
\begin{equation}
\lambda_T= {1\over3}\sqrt{{|\vartheta|\over[q +(2\zeta\vartheta/3H^2)]H}}\,\lambda_H\,.  \label{BTS}
\end{equation}
As in the sub-Hubble case (ii) discuss before, the deceleration parameter turns negative on lengths smaller that $\lambda_T$. Compared to the FRW studies (e.g.~see Eq.~(22) in~\cite{T4} or (\ref{FRWTS}) here), the background anisotropy has added the shear-related term in the denominator of the above. Put another way, in the absence of anisotropy, $\zeta=0$ and expression (\ref{BTS}) reduces to its counterpart obtained in tilted Einstein-de Sitter and generalised FRW universes. There, as well as here, the transition length is a fraction of the Hubble scale and closely analogous to the Jeans length~\cite{T3}.\footnote{The Jeans length ($\lambda_J\sim c_s\lambda_H$, with $c_s$ representing the sound speed -- e.g.~see~\cite{TCM,EMM}) marks the scale below which weak pressure perturbations dominate over the background gravitational pull. In analogy, the transition length marks the scale below which weak peculiar-velocity perturbations ``dominate'' over the background expansion.} Here, the anisotropy of the Bianchi background will increase $\lambda_T$ when $\zeta>0$, or decrease it if $\zeta<0$, though the effect is weak in either case (in realistically anisotropic models with $|\zeta|/H\ll1$).

According to the current observations, the anisotropy of the universe is small (if any). Therefore, in all realistic scenarios we may assume that $|\zeta|/H\ll1$. Then, on sub-horizon scales, expressions (\ref{ltq4}) and (\ref{BTS}) are replaced by their tilted Einstein-de Sitter counterparts, namely by Eqs.~(\ref{ltq6}) and (\ref{FRWTS}) respectively.\footnote{Just like in the tilted almost-FRW universes (see~\cite{T4} for further discussion and technical details), including pressure and spatial curvature to tilted almost-Bianchi cosmologies, with a physically reasonable degree of anisotropy, will make no difference to Eqs.~(\ref{ltq6}) and (\ref{FRWTS}) on scales well inside the Hubble horizon.} Applying these relations to the observational data, one can estimate the associated values of the deceleration parameter and of the corresponding transition scale measured by observers living inside the reported bulk peculiar flows (see Table~\ref{tab2}).

\begin{table}
\caption{Representative estimates of the deceleration parameter ($\tilde{q}$ -- see Eq.~(\ref{ltq6})), as measured by observers residing within some of the bulk flows reported in~\cite{CMSS,Maetal,ND} and~\cite{Wetal}, assuming that the latter are slightly contracting. The last column provides the associated transition lengths ($\lambda_T$ -- see Eq.~(\ref{FRWTS})), marking the $\tilde{q}=0$ threshold. Note that the scale dependence seen in (\ref{ltq6}), according to which (for comparable values of the bulk velocity) the peculiar-motion effect on $\tilde{q}$ increases on smaller scales and decreases as we move out to larger lengths, is clearly manifested in this Table. Also, the (mean) transition length exceeds the bulk-flow scale in all cases. The variations in the values of $\tilde{q}$ and $\lambda_T$ reflect the differences between the reported bulk peculiar velocities, which have not reached a consensus yet. The surveys are listed in chronological order, with the earlier ones tending to have larger error bars. Finally, $\lambda$ and $\lambda_T$ are measured in Mpc, $\langle v\rangle$ in km/sec, while we have set $q=1/2$, $\vartheta=\langle\vartheta\rangle\simeq -\langle v\rangle/\lambda$ and $H\simeq70$~km/sec\,Mpc today.}
\begin{center}\begin{tabular}{cccccccc}
\hline \hline & \hspace{-30pt} Survey & \hspace{-12pt} $\lambda$ & \hspace{-12pt} $\langle v\rangle$ & \hspace{-10pt} $\tilde{q}$ & \hspace{-10pt} $\lambda_T$ &\\ \hline \hline & $\begin{array}{c} \hspace{-20pt} {\rm Ma,\, et al\; (2011)} \\ \hspace{-20pt} {\rm Colin,\,et\,al\;(2011)} \\ \hspace{-20pt} {\rm Nusser\,\&\,Davis\;(2011)} \\ \hspace{-20pt} {\rm Watkins,\,et\,al\;(2023)} \\ \hspace{-20pt} {\rm Watkins,\,et\,al\;(2023)} \end{array}$ & $\begin{array}{c} \hspace{-10pt} 150 \\ \hspace{-10pt} 250 \\ \hspace{-10pt} 150 \\ \hspace{-10pt} 200 \\ \hspace{-10pt} 250 \end{array}$ & $\begin{array}{c} \hspace{-8pt} 340^{_{+130}}_{_{-130}} \\ \hspace{-8pt} 260^{_{+190}}_{_{-150}} \\ \hspace{-7pt} 250^{_{+40}}_{_{-40}} \\ \hspace{-7pt} 395^{_{+30}}_{_{-30}} \\ \hspace{-7pt} 430^{_{+40}}_{_{-40}} \end{array}$ & $\begin{array}{c} \hspace{-7pt} -3.90^{_{-1.70}}_{_{+1.70}} \\ \hspace{-7pt} -0.25^{_{-0.75}}_{_{+0.20}} \\ \hspace{-7pt} -2.75^{_{-0.50}}_{_{+0.50}} \\ \hspace{-7pt} -1.70^{_{-0.15}}_{_{+0.15}} \\ \hspace{-7pt} -0.70^{_{-0.15}}_{_{+0.15}} \end{array}$ & $\begin{array}{c} \hspace{-7pt} 440^{_{+80}}_{_{-80}} \\ \hspace{-8pt} 300^{_{+95}}_{_{-105}} \\ \hspace{-7pt} 380^{_{+30}}_{_{-30}} \\ \hspace{-7pt} 415^{_{+15}}_{_{-15}} \\ \hspace{-7pt} 390^{_{+15}}_{_{-15}} \end{array}$\\ [2.5truemm] \hline \hline
\end{tabular}\end{center}\label{tab2}\vspace{-10pt}
\end{table}

\section{Discussion}\label{sD}
Imagine a passenger sitting in the back of a car that drives in a wide motorway, with all the cars moving at the same speed. Suppose also that, without the passenger realising it, their car slows down a little. Everyday experience then tells us that this can easily mislead the unsuspecting passenger to believe that it is the rest of the vehicles that have accelerated away. Is it then possible that something analogous can also happen to observers residing in a slightly contacting bulk peculiar flow? \textit{Is it possible that the unsuspecting observers may misinterpret their slower local expansion rate as recent global acceleration of the surrounding universe?}

The theoretical possibility that the accelerated expansion of the universe is a mere illusion triggered by our peculiar motion was originally raised in~\cite{T3,T4}. That scenario was later revisited and refined, by applying relativistic perturbation theory to a tilted Einstein-de Sitter model~\cite{TK,T3} and subsequently extended to include all three FRW universes~\cite{T4}. The latter study found that neither the pressure nor the spatial curvature could significantly change the relative-motion effects on the local deceleration parameter. The only exception was in the case of Friedmann universes with closed spatial sections, but with unrealistically high positive curvature~\cite{T4}. In all these studies, observer's living inside bulk peculiar flows, like those reported in~\cite{WFH}-\cite{Wetal} for example, could be easily misled to believe that their universe was accelerating. This could happen if the bulk flows in question were locally contracting, which is what our local peculiar-velocity field may be actually doing~\cite{PGC}. Then, the unsuspecting observers could misinterpret their local contraction as acceleration of the surrounding universe, just like passengers in a car that slows down may think that the vehicles around them have accelerated away.

Here, the work of~\cite{T3,T4} has been extended to include the anisotropic Bianchi-type universes. Our aim was to investigate whether the effects of anisotropy could significantly change the tilted Einstein-de Sitter picture. We found that, in principle, the shear anisotropy of the Bianchi spacetimes could modify the relative-motion effects on the locally measured deceleration parameter, either by enhancing or by reducing them. Nevertheless, unless the universe has unrealistically high anisotropy, the results of the Einstein-de Sitter scenario reported in~\cite{TK,T3} remain unchanged. This is especially true on subhorizon scales, where the crucial scale-dependence (${\lambda_H/\lambda}$) of the peculiar-motion effects reported in~\cite{TK}-\cite{T4} is identical to the one seen in Eq.~(\ref{ltq4}) here. Put another way, as with the pressure and the spatial curvature, the anisotropy does not seem capable of altering the Einstein-de Sitter picture. Slightly contracting bulk peculiar motions, like those recently reported in~\cite{PGC}, can create the illusion of global acceleration by causing an apparent change in the sign of the local deceleration parameter (see Table~\ref{tab2} above). The effect is local, but the \textit{transition length}, where the deceleration parameter appears to turn negative, is large enough (of the order of few to several hundred Mpc) to make it look like a recent global event~\cite{TK}-\cite{T4}.

Both (\ref{ltq6}) and (\ref{FRWTS}), as well as the resulting $\tilde{q}$-profile seen in Figs.~\ref{fig:tq-1} and \ref{fig:tq-2}, hold in all physically realistic FRW and Bianchi universes, provided the bulk flow in question is (slightly) contracting. Hence, qualitatively speaking, the physical mechanism behind the relative-motion effects on the locally measured deceleration parameter seems generic, as it appears independent of the background cosmology. In fact, the close analogies between the transition length ($\lambda_T$) and the Jeans length ($\lambda_J$ -- see footnote~10 here and also~\cite{T3}) suggest that the former is as generic to peculiar-velocity perturbations as the letter is to density perturbations. Also note that the scale dependence of the linear relative-motion effects on $\tilde{q}$, raises the intriguing possibility that they may even dominate during the nonlinear phase of structure formation. After all, on the nonlinear scales, typically those with $\lambda\ll100$~Mpc, the ratio $({\lambda_H/\lambda})^2$ takes very large values.

The tilted-universe scenario does not appeal to exotic matter or/and to new physics to address the ongoing question of the universal acceleration. There is neither dark energy nor a cosmological constant, and the scenario operates within standard general relativity and conventional cosmology. Instead, the recent accelerated expansion of the universe is explained as a local illusion triggered by the peculiar motion of our galaxy. The latter is an inevitable byproduct of structure-formation and a hard observational fact, all of which provide a solid physical motivation to the tilted scenario. Moreover, since peculiar motions were only recently added to the kinematics of our universe, the tilted model faces no ``coincidence problem'' either.

In addition to its simplicity and physical motivation, the proposed scenario also makes three specific predictions that can be tested against the observations, like the Pantheon+ data-set and the upcoming LSST-DA survey. The first is the predicted profile of the deceleration parameter, which takes positive values on sufficiently high redshifts and naturally turns negative nearby (see Figs.~\ref{fig:tq-1} and \ref{fig:tq-2} in \S~\ref{ssOEtq} above). The second is the trademark signature of relative motion, namely an apparent (Doppler-like) dipole in the observed distribution of the deceleration parameter. To the bulk-flow observers the universe should appear to accelerate faster towards a certain direction in the sky and equally slower along the antipodal~\cite{T4}. Dipolar anisotropies of this sort in the universal acceleration have been reported in the literature, with the first claim (to the best of our knowledge) made in~\cite{CL-B}. It was not until~\cite{CMRS} and more recently in~\cite{Cetal}, however, that these anisotropies where attributed to our galaxy's peculiar motion. Nevertheless, the current limitations of the observational sample still cloud the statistical significance of the findings~\cite{BAP}.\footnote{A dipole in the distribution of the deceleration parameter, due to the observer's bulk flow, could imply the same for the Hubble parameter as well~\cite{Metal,Letal}, since the former is the time derivative of the latter. For the same reason, the magnitudes and the directions of the two dipoles do not need to coincide.} Finally, a key third prediction of the tilted-universe scenario is that the magnitude of the $\tilde{q}$-dipole should decrease with redshift, given that peculiar velocities also fade away on progressively larger scales.\\

\textbf{Acknowledgements:} The authors would like to thank Kerkyra Asvesta for her help with Figs.~2 and~3. CGT also acknowledges financial support from the Hellenic Foundation for Research and Innovation (H.F.R.I.), under the ``First Call for H.F.R.I. Research Projects to support Faculty members and Researchers and the procurement of high-cost research equipment Grant'' (Project No. 789).

\end{document}